\documentclass{aastex63}
\usepackage{comment}
\usepackage{mathrsfs}
\usepackage{subfigure,wrapfig}

\revised{\today}
\submitjournal{ApJ}

\shorttitle{Turbulently-Driven Helium Detonation Initiation}
\shortauthors{Casabona et al.}

\begin{document}

\title{Turbulently-Driven Detonation Initiation in Electron-Degenerate Matter with Helium}

\correspondingauthor{Gabriel Casabona}
\email{casabona@u.northwestern.edu }

\author[0000-0002-0786-7307]{GABRIEL O. CASABONA}
\affiliation{Northwestern University \\
2145 Sheridan Road \\
Evanston, IL 60208-3112 }

\author[0000-0001-8077-7255]{Robert T. Fisher}
\affiliation{University of Massachusetts Dartmouth \\
285 Old Westport Road \\
Dartmouth, MA 02747-2300}

\begin{abstract}

Type Ia supernovae (SNe Ia) are standardizable cosmological candles which led to the discovery of the accelerating universe. However, the physics of how white dwarfs (WDs) explode and lead to SNe Ia is still poorly understood. The initiation of the detonation front which rapidly disrupts the WD is a crucial element of the puzzle, and global 3D simulations of SNe Ia cannot resolve the requisite length scales to capture detonation initiation. In this work, we elucidate a theoretical criterion for detonation initiation in the distributed burning regime. We test this criterion against local 3D driven turbulent hydrodynamical simulations within electron-degenerate WD matter consisting initially of pure helium.
We demonstrate a novel pathway for detonation, in which strong turbulent dissipation rapidly heats the helium, and forms carbon nuclei sufficient to lead to a detonation through accelerated burning via $\alpha$ captures. Simulations of strongly-driven turbulent conditions 
lead to detonations at a mean density of  $10^6$ g cm$^{-3}$ and mean temperature of $1.4 - 1.8 \times 10^9$ K, but fail to detonate  at a lower density of $10^5$ g cm$^{-3}$, in excellent agreement with theoretical predictions.

\end{abstract}


\section{Introduction} \label{sec:intro}

Type Ia supernovae (SNe Ia) are the thermonuclear explosions of carbon-oxygen white dwarfs (WDs) in binary systems, in which the companion star accretes matter onto the WD \citep {maozetal14}. 
Their use as standardizable candles led to the discovery of the accelerated expansion of the universe \citep {pankey62, Phillips, Schmidt_1998, Riess, Perlmutter_1999}. SNe Ia are also prominent sources of cosmic rays \citep [e.g.][]{sanoetal19} and iron group elements \citep [e.g.][]{kobayashietal20}.

Although it is known that normal SNe Ia are the thermonuclear explosions of carbon-oxygen WDs \citep {bloometal11}, their detonation mechanism remains unclear. Motivation for this current paper comes from a consideration of the role of helium in leading models for SNe Ia.
In one possible scenario, referred to as the double detonation scenario, a C/O WD accretes mass from a helium donor \citep {taam80a, taam80b, nomoto82, woosleytaamweaver86}. The helium layer eventually becomes convectively unstable, leading to the formation of an  individual hot spot and detonation \citep {jacobsetal16}. The helium detonation sends shock waves inwards through the C/O core. Multidimensional simulations have revealed a ``scissors'' mechanism, in which the shock waves meet within the carbon- and oxygen-enriched mixed transition layer from the core to the helium shell \citep {gronowetal20}. At the point where the shock waves meet, carbon detonation is initiated at an off-center location.

Another promising explosion mechanism traces back to the fact that C/O WDs have a relatively thin helium shell around them, resulting from stellar evolution \citep {Giammichele}. During the merger of a  binary system of two C/O WDs, the helium of the secondary WD will accrete and mix with the helium layer of the primary WD, and may lead to first a detonation of the helium layer and then the primary \citep {guillochonetal10}. 
Recent observations of the surviving ex-companion WDs from these dynamically driven double-degenerate double-detonation (or D$^6$) SNe Ia from Gaia supports this model as the origin of at least some SNe Ia \citep {ShenGaia, elbadryetal23}.

Previous work  constrained the temperature, density, and critical length scales needed for helium detonation based upon the Zel'dovich gradient mechanism within a shallow temperature gradient ramp on a static,  laminar background \citep {zeldovich70, zeldovich80, blinnikovkhokhlov86, Holcomb_2013}. These models were computed by using one-dimensional laminar hydrodynamic simulations with a variety of initial conditions, motivated by typical WD conditions expected in various type Ia explosion scenarios. 
However, the actual hydrodynamic conditions expected in both the double detonation and D$^6$ scenarios are at extremely high Reynolds and Karlovitz numbers, ${\rm Re}  = L v_0 / \eta \sim 10^{15} - 10^{16}$ and ${\rm Ka} = \left[ (v_0 / S_l)^3 l / L \right]^{1/2} \sim 10^5 - 10^9$, where $v_0$ is the turbulent RMS velocity on the scale $L$, $S_l$ is the laminar speed of a flame of laminar width $l$ \citep {timmeswoosley92}, and $\eta$ is the kinematic viscosity  \citep {nandkumarpethick84}. Under such conditions with ${\rm Re} >> 1$ and ${\rm Ka} >> 1$, helium burning will occur in the distributed burning regime, with the local laminar flame surface completely disrupted by the influence of turbulence, and spread over the turbulent integral scale \citep {ropkehillebrandt05}.  Previous work has begun to reveal how turbulence influences detonation initiation in electron degenerate conditions \citep{woosley07, aspdenetal10, poludnenkoetal11, fennplewa17, poludnenkoetal19, Fisher, brookerplewafenn21, zenatifisher23}. In particular, using both  local three dimensional hydrodynamics simulations and analytic calculations, this body of work demonstrates how turbulent, electron degenerate matter may realistically detonate in both the flamelet and in the distributed burning regime. We refer to this detonation mechanism as the \textit{turbulently-driven detonation mechanism}. 

In this current work, we explore the interplay of turbulence and nuclear burning in helium environments. We focus carefully on the simplest cases of pure helium fuel,
and seek to understand the fundamental physics of turbulent detonation initiation in the distributed regime for a simplified nuclear network, using both analytic criteria as well as three-dimensional numerical simulations. We find that vigorous turbulent burning under conditions expected in helium-accreting WD scenarios naturally gives rise to turbulent heating and the rapid nucleosynthesis of seed carbon nuclei, even upon an initially pure helium background.  The combined effects of strong turbulent dissipation and $\alpha$ captures on to seed carbon and oxygen nuclei can then promote the detonation of helium under broader thermodynamic conditions than predicted for laminar flows.

\section{Distributed Nuclear Helium Burning and Detonation Initiation}\label{sec:floats}

To gain insight into the basic physics of turbulent nuclear helium burning, we first consider an analytic detonation criterion within the turbulent nuclear helium burning layer. We employ a mean field or ``one zone" model, neglecting turbulent fluctuations in the thermodynamic state of the gas, and computing all other quantities dependent upon the thermodynamic state  (such as sound speed or burning rate) at their  mean values.\footnote{A justification of the mean field assumption is provided in the discussion Section \ref {sec:discussion}.} A general condition for detonation initiation in the mean field approximation is that the rate of nuclear energy release within the flame is faster than the sound-crossing time across the flame, 

\begin {equation}
\epsilon_{\rm nuc} \ge {e_{\rm int} \over \tau_s} =  {1 \over \Gamma_c \left (\Gamma_e - 1\right)} {c_s^3 \over L}
\label {detcond}
\end {equation}
where $e_{\rm int}$ is the specific internal energy of the gas, and $\tau_s = L / c_s$ is the sound-crossing time across the flame length $L$ \citep {poludnenkoetal19}.  The second equality follows immediately from the definitions of the adiabatic exponents $\Gamma_e = P (\rho e_{\rm int} )^{-1} + 1$, and  $\Gamma_c = c_s^2 \rho  P^{-1}$, where $\rho$ is the mass density, $P$ is the pressure, and $c_s$ is the sound speed. \citet {poludnenkoetal11} and \citet {poludnenkoetal19} show that for both turbulent as well as distributed flames, equation \ref {detcond} is equivalent to the condition that the turbulent flame speed $S_T$ exceeds the Chapman-Jouguet deflagration speed $S_{\rm CJ}$ for an ideal gas equation of state: $S_T > S_{\rm CJ}$.

Turbulent nuclear burning in the helium surface layers of the white dwarf occurs in the distributed burning regime, in which the flame is disrupted and spread over an integral scale $L$, on which it achieves an integral scale RMS turbulent velocity $v_0$. Rearranging equation \ref {detcond} for a turbulent distributed flame in which the velocity statistics are  Kolmogorov, we find

\begin {equation}
\epsilon_{\rm nuc} \ge {1 \over C_K \Gamma_c \left (\Gamma_e - 1\right)} {\epsilon_{\rm turb} \over \mathcal {M}_0^3 }
\label {distburn}
\end {equation}
Here, $\epsilon_{\rm turb} = C_K v_0^3 / L$ is the specific turbulent dissipation rate, with $v_0$ the RMS turbulent velocity on the integral scale $L$, and $C_K$ a universal dimensionless constant known as Kolmogorov's constant, which is approximately 0.5 in the limit of large Reynolds numbers \citep {kanedaetal03}.  $\mathcal {M}_0$ is the integral scale turbulent RMS Mach number of the flame, $\mathcal {M}_0 = v_0 / c_s$. 

Thus, the intuitive criterion that the burning occurs in the nuclear-dominated regime ($\epsilon_{\rm nuc} / {\epsilon_{\rm turb} } \geq 1$) is necessary but not sufficient for detonation. Precisely as one might physically expect, equation \ref {distburn} shows that the nuclear burning rate must exceed the nominal nuclear-dominated criterion of $\epsilon_{\rm nuc} \geq \epsilon_{\rm turb}$ by a dimensionless factor $C_K^{-1} \Gamma_c^{-1} (\Gamma_e - 1)^{-1} \mathcal {M}_0^{-3} $, which becomes increasingly more stringent with increasing temperature. The monotonic increase of the detonation initiation condition with respect to temperature arises very simply, both because the specific internal energy of the gas increases, and the sound crossing time decreases, as the electron degeneracy is gradually lifted in equation \ref {detcond}.

We further note that, neglecting turbulent intermittency, both the turbulent dissipation $\epsilon_{\rm turb}$ (which is a spatial invariant throughout the inertial range for Kolmogorov velocity scaling), as well as the nuclear burning rate $\epsilon_{\rm nuc}$ are independent of scale throughout the inertial range in the mean field approximation. The only scale-dependent factor which remains in  equation \ref {distburn} is the integral scale turbulent RMS Mach number ${\cal M}_0$. We can write the RMS Mach number on the scale $r$ as ${\cal M}_r = {\cal M}_0 (r / L)^{1/3}$ in the inertial range, and use equation  \ref {distburn} to express a minimum critical length for detonation initiation in distributed burning:

\begin {equation}
r_{\rm crit} \ge {1 \over C_K \Gamma_c \left (\Gamma_e - 1\right)} {\epsilon_{\rm turb} \over \epsilon_{\rm nuc} } L 
\label {rcrit}
\end {equation}
In other words, when the equality in the nuclear burning condition given in equation \ref {distburn} is met, the minimum detonation size is the integral scale $L$. For a fixed turbulent background, as the the nuclear energy generation rate is increased beyond this value, the critical length as derived in equation \ref {rcrit} decreases. Conversely, if the nuclear burning rate fails to meet Equation \ref {distburn}, the inferred critical length scale nominally exceeds the integral scale $L$, and no detonation will be realized within a volume extending across the integral scale.  



\begin{wrapfigure}{l}{0.5\textwidth} 
    \centering
    \includegraphics[width=0.45\textwidth]{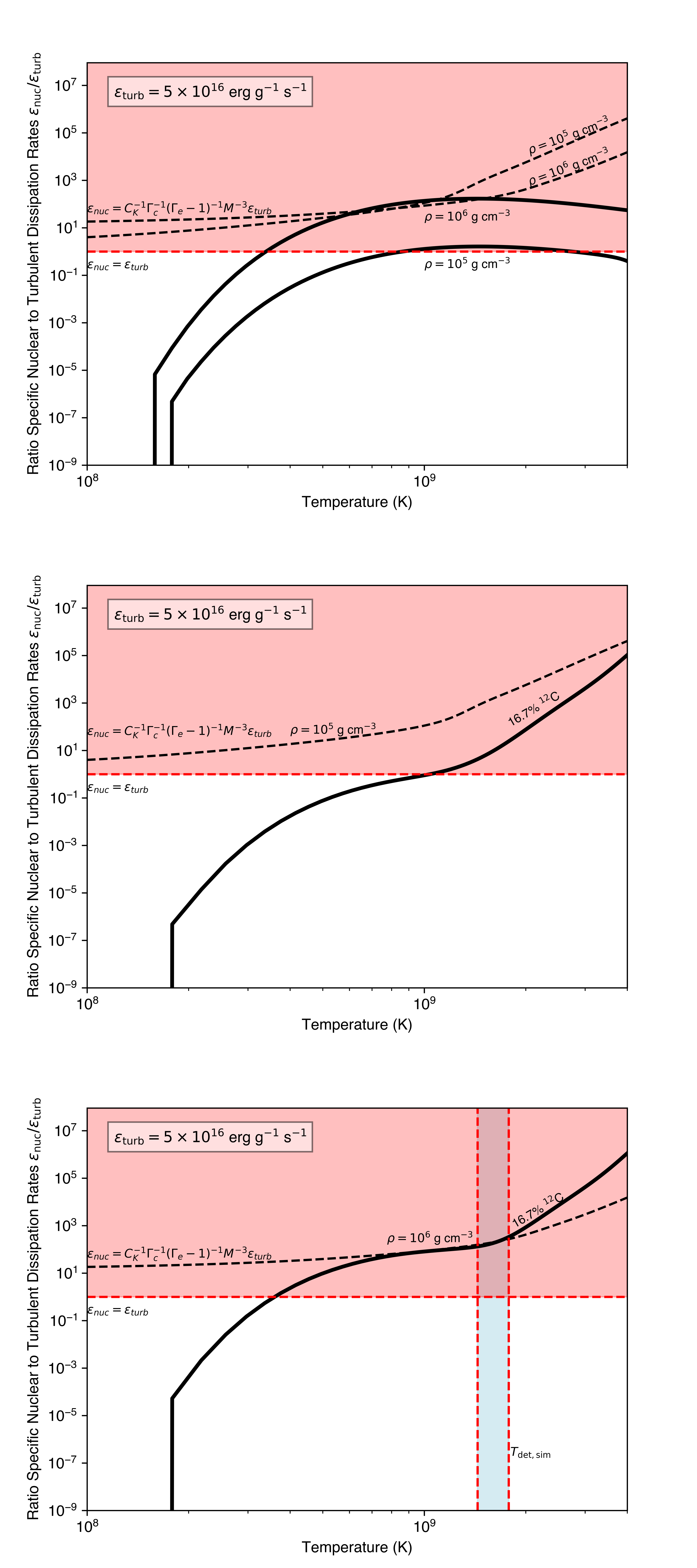}
    
\caption{Analytic curves showing the ratio of the specific nuclear energy rates to the specific turbulent heating rate for the numerical experiments undertaken here: top) pure helium,  middle) 83.3\% $^{4}$He with  16.7\% $^{12}$C seed nuclei at a density of  10$^5$ g cm$^{-3}$, and bottom) the same composition at 10$^6$ g cm$^{-3}$. 
}
\label {fig:he_analytic}

\end{wrapfigure}

We next compare nuclear burning with turbulent heating in the simplest case of pure helium. Figure \ref {fig:he_analytic} shows the ratio of  the specific nuclear burning rate  $\epsilon_{\rm nuc}$ to the specific turbulent dissipation rate $\epsilon_{\rm turb} = C_K v_0^3 / L$, where $v_0$ is the RMS turbulent velocity on the scale $L$, and $C_K \simeq 0.5$ is the universal Kolmogorov constant introduced earlier. Nuclear burning becomes the dominant heating mechanism when $\epsilon_{\rm nuc}/\epsilon_{\rm turb} \ge 1 $. Both neutrino losses \citep {itohetal96} as well as electron screening \citep {wallaceetal82} are taken into account for the burning curves. The adiabatic indices $\Gamma_c$ and $\Gamma_e$ are also numerically calculated for the critical detonation condition, equation \ref {distburn}, using the Helmholtz equation of state \citep {timmesswesty00}. The plot shows the dependence of the ratio $\epsilon_{\rm nuc}/\epsilon_{\rm turb}$ upon the temperature, holding the turbulence parameters fixed.  We adopt a fiducial value of $\epsilon_{\rm turb} = 5 \times 10^{16}$ erg g$^{-1}$ s$^{-1}$, corresponding to $v_0 = 10^3$ km s$^{-1}$ on the scale $L = 100$ km, which is motivated by a typical accretion stream velocity  in helium-ignited double-degenerate systems, and the corresponding Kelvin-Helmholtz instability driving scale  \citep{guillochonetal10, ShenGaia}. The analytic curves on the top plot consider pure helium at densities of  $10^5$ and $10^6$ g cm$^{-3}$, representative of varying depths within the surface helium of massive WDs thought to be responsible for double detonation and D$^6$ SNe Ia.

A critical temperature, ranging from $\sim 1 - 2 \times 10^8$ K depending on density, is required to offset the neutrino losses and ignite unstable He burning. Unlike $^{12}$C-$^{12}$C burning, which continues to rise with temperature, the triple-$\alpha$ reaction exhibits a maximum value close to $T = 1.5 \times 10^9$ K.  Fundamentally, this maximal specific nuclear energy rate  occurs because triple-$\alpha$ hinges upon the abundance of the resonant production of $^{8}$Be, which peaks at approximately $7 \times 10^8$ K in equilibrium with $\alpha$ + $\alpha$ \citep {clayton68}.

Consequently, simply increasing the temperature of a pure helium layer does not lead to an increase in the nuclear burning rate beyond $T = 1.5 \times 10^9$ K; one also requires an increase of density to enhance the burning. As a result, the fundamental nuclear physics of the triple-$\alpha$ reaction requires that there is a critical density needed to enter into the nuclear-dominated heating regime for turbulently-driven pure helium, above approximately $10^5$ g cm$^{-3}$ for typical double-degenerate merger systems. 

Furthermore, as is evident from the top of Figure \ref {fig:he_analytic}, even though the higher densities enter into the nuclear-dominated heating regime for some temperatures,  pure He burning fails to meet Equation  \ref {distburn} for all but a thin sliver of temperatures at densities of $\sim  10^6$ g cm$^{-3}$. {\it However, critically, the assumption of pure helium needs to be loosened, because strong turbulent dissipation heats the helium and gives rise to seed nuclei which rapidly accelerate the burning.}

Accordingly, in the middle  of Figure \ref  {fig:he_analytic}, we consider the lower density case of $10^5$ g cm$^{-3}$, we show an admixture of  $^{4}$He with seed $^{12}$C nuclei. We choose 16.7\% $^{12}$C and 83.3\% $^4$He, in order to facilitate comparison against the numerical simulations to be presented in Section \ref {sec:results}.
Even the introduction of 16.7\% $^{12}$C is still insufficient to meet the detonation initiation criterion at any temperature below $3 \times 10^9$ K  at this lower density, with its lower specific nuclear burning rate.

Finally, on the bottom of Figure \ref  {fig:he_analytic}, we show a higher mean density of $10^6$ g cm$^{-3}$, and the same composition as the middle.  Here, the admixture of $^{12}$C nuclei sufficiently enhances the burning rate above $10^9$ K due to $\alpha$ captures, and leads to a range of temperatures satisfying the detonation condition above $1.5 \times 10^9$ K. The actual range of temperatures of the detonation obtained in the numerical simulation is shown in the blue region.

In order to test these analytic insights further, and to explore the prospects for helium detonation in a realistic turbulent flow, we next incorporate helium into three-dimensional hydrodynamical simulations.  We have carried out three-dimensional hydrodynamic simulations, and have explored the regimes for helium detonability under the influence of the turbulently-driven detonation mechanism.

\section{Simulation Methodology}\label{sec:method}

FLASH 4.0.1, a multiphysics multiscale code designed to simulate astrophysical and high-energy density plasmas, was used to undertake driven three-dimensional turbulence simulations with nuclear burning \citep {fryxelletal00}.
The hydrodynamics was solved using the split piecewise parabolic method. The Helmholtz equation of state was used to incorporate the contributions from nuclei, electrons, blackbody photons, electron-degeneracy, and an arbitrary degree of special relativity  \citep{timmesswesty00}. Since we are primarily interested in helium burning and the burning associated with light elements including carbon and oxygen, in this paper we employ the  19-isotope network used in \citet{Weaver} and \citet{Timmes}, which includes $\alpha$ captures and neutrino losses.

In order to explore the analytic predictions developed in the last section in a more realistic setting, the density parameter space was explored using fully time-dependent turbulent hydrodynamical simulations. 
Two simulations were conducted, with initial densities  set to $10^5$ (LowDen) and $10^6$ (HighDen) g cm$^{-3}$. Each run employed fully-periodic uniform grid Cartesian geometries of $512^3$ cells; or equivalently, linear spatial resolution of 195 m. In order to explore the analytic predictions developed in the last section in a more realistic setting, the parameter space of density and nuclear composition was explored. Initial densities were set to $10^5$ (LowDen) and $10^6$ (HighDen) g cm$^{-3}$. The initial helium abundance was set at 100\% in both cases.

The initial temperature in all simulations began prior to driving at $10^8$ K; turbulent dissipation causes the temperature to increase as the simulation proceeds, and reaches $\sim 3 \times 10^8$ K by the end of driving. Because of this physical effect of turbulent heating, each model can be understood to sweep through a range of background temperatures at fixed density (due to periodic boundaries), beginning at $3 \times 10^8$ K.

Simulations begin with the fluid having zero velocity, and uniform density, composition, and temperature.
A large-scale stochastic forcing routine is then used to increase the turbulent velocity of the fluid. Each simulation runs until the RMS velocities and global enstrophies reach  approximately stable values, indicating that the runs have reached steady-state turbulence.  An RMS velocity of $v_0 = 1.25 \times 10^8$ cm s$^{-1}$ was established, approximately the same as the large-scale shear velocity induced by accretion in double-degenerate systems \citep{ShenGaia}.  Once a steady-state RMS turbulent velocity is achieved, the simulations are restarted with nuclear burning turned on, which we define as $t = 0$, and are continued to determine whether detonation will occur. Detonation arises as  a  supersonic burning of the helium fuel accelerated by $^{12}$C seed nuclei, whose onset is quantified according to two separate criteria as explained below.

\section{Simulation Results}\label{sec:results}

We present in detail the HighDen run. Figure \ref {fig:c12_enuc} shows slice plots for the HighDen run at the time when the detonation has initiated, with the box centered about the point of maximum temperature. The temperature hot spot also coincides with the maximum carbon abundance. Each slice plot has an inset which is zoomed in on the hotspot.
 
\begin{figure}[ht!  ]

\includegraphics[width=1.0\textwidth]{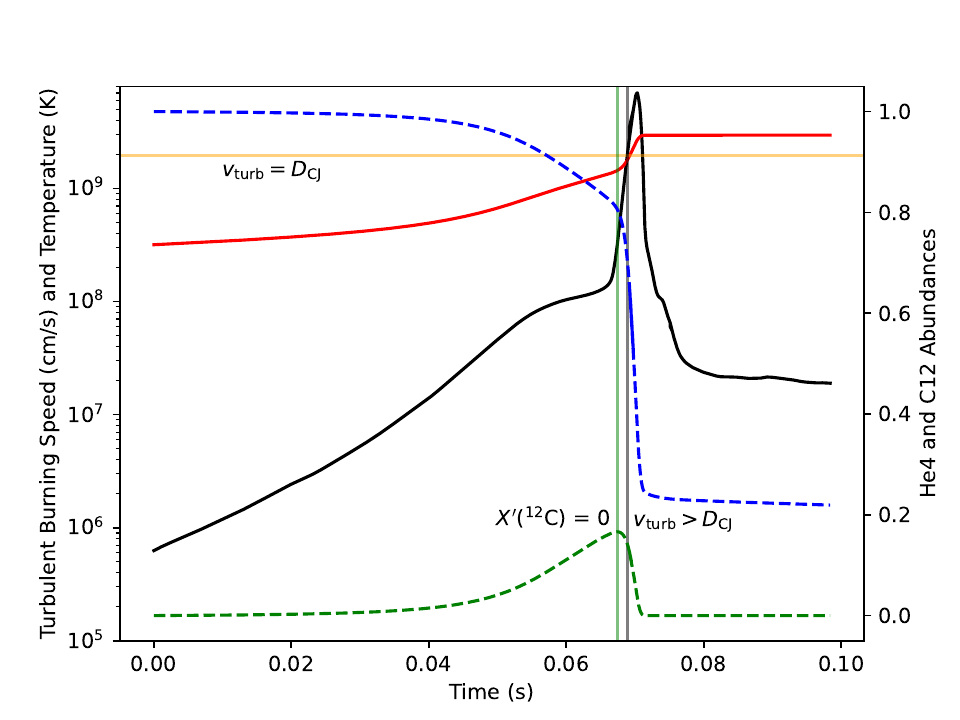}

\caption{A plot showing the effective turbulent burning speed in black in cm/s, the mean temperature in red in K (both on left axis), and the global abundances of both $^{4}$He (dashed blue) and $^{12}$C (dashed green) (on right axis)  in the  HighDen model. The horizontal orange line marks the Chapman-Jouguet detonation velocity calculated from the final ash state. The vertical black line shows the onset of detonation in this model as determined by when $v_{\rm turb} > D_{\rm CJ}$. The vertical green  line indicates the time at which the $^{12}$C abundance is maximal.}
\label {fig:vturb}

\end{figure}

\begin{figure}[ht!  ]

\includegraphics[width=1.0\textwidth]{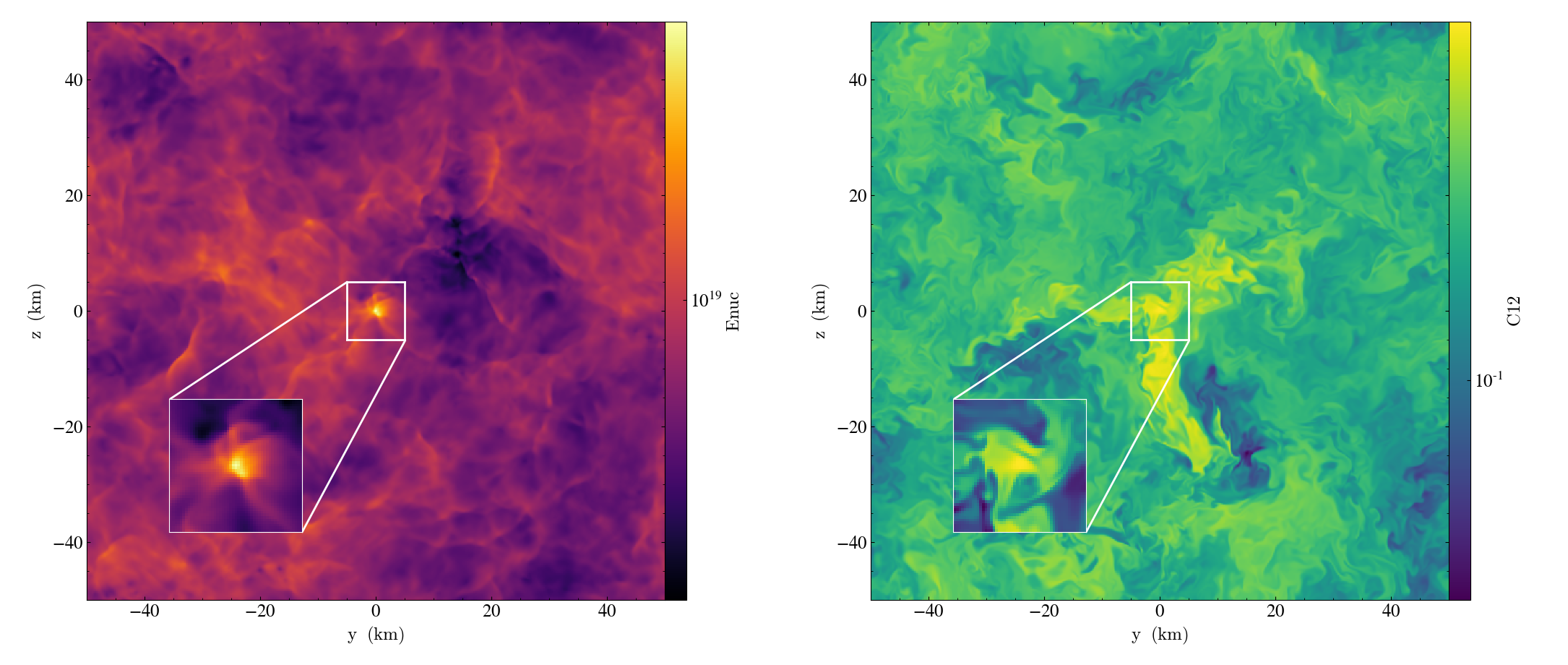}

\caption{Slice plots of specific nuclear energy generation rate and abundance ratio of C from HighDen run, through the x-axis in the $z$-$y$ plane. Each is centered about the hot spot just before leading into a detonation.}
\label {fig:c12_enuc}

\end{figure}

Figure \ref {fig:vturb} shows the effective turbulent burning speed as well as the global $^4$He and $^{12}$C mass abundance fractions as a function of time. The effective distributed turbulent burning speed across the simulation domain is defined as $v_{\rm turb} \equiv L / t_{\rm burn}$, where $t_{\rm burn}$ is the $^{4}$He burning timescale, $t_{\rm burn} = X (^4 {\rm He} ) |d X (^4 {\rm He} ) / dt|^{-1} $. At $t = 0.12$ s, the effective turbulent flame speed is $6 \times 10^5$ cm/s. Turbulent dissipation slowly increases internal energy within the box, which also increases the temperature from its initial value of $3.2 \times 10^8$ K as the electron degeneracy is lifted, and the nuclear burning rate for 3$\alpha$ rises in this same temperature range, as shown in Figure \ref  {fig:he_analytic}. By 0.18 s, the turbulent flame speed eventually reaches a brief steady state in near-equality with the turbulent RMS velocity around $10^8$ cm/s. \citet {damkohler40} showed that $v_{\rm turb} \sim v_0$, often known as Damkohler scaling,  could be predicted from the kinematic interaction of the flame surface with turbulence when the turbulent RMS velocity greatly exceeds the laminar flame speed. Here, however, the balance is struck rather accidentally owing to the plateau of 3$\alpha$ above $7 \times 10^8$ K. At the same time, however, the abundance of $^{12}$C seed nuclei builds up, rising from its initial value of 0\% to 16.7\%. After the temperature has risen above $10^9$ K, $\alpha$ captures onto seed $^{12}$C nuclei dominate over 3$\alpha$ in  the nuclear energy generation rate. Soon thereafter, as the temperature hits $1.5 \times 10^9$ K, a very sharp increase in the nuclear burning rate takes place, signaling the onset of detonation.  $^{16}$O abundances also rise slightly before detonation to $2.8 \times 10^{-4}$.   $^{16}$O is a more effective seed nucleus for $\alpha$ captures than $^{12}$C, thanks to the 5.8 MeV $J^{\pi} = 1^-$ state in $^{20}$Ne, which facilitates the resonant interaction $^{16}$O $(\alpha, \gamma)^{20}$Ne at temperatures $T\sim 8 \times 10^8$ K and above \citep {clayton68}. However, the $\alpha$ capture process is relatively inefficient in converting seed  $^{12}$C to $^{16}$O in this run. Specifically, the low ratio of $^{16}$O/$^{12}$C demonstrates that the $\alpha$ capture chain is terminated primarily at $^{12}$C by the onset of detonation in this model.

The detonation onset is analyzed using two different criteria. The first is a purely phenomenological determination based upon the time of maximal abundance of the seed nucleus $^{12}$C. This first criterion makes no assumptions of the mechanism or the physics underlying the detonation, other than that $\alpha$ captures onto $^{12}$C dominate the 3$\alpha$ process. The second criterion compares the effective flame speed against the Chapman-Jouguet detonation velocity $D_{\rm CJ}$. Combustion theory shows that the minimum speed of a steady-state detonation is the Chapman-Jouguet detonation velocity \citep {lee08}. We compute $D_{\rm CJ}$ at the final ash composition state and temperature of $T = 3 \times 10^9$ K using the Helmholtz equation of state, and overplot it as a solid orange horizontal line in Figure  \ref  {fig:vturb}. Both detonation criterion closely coincide, yielding detonation temperatures of $1.44 \times 10^9$ K and $1.78 \times 10^9$ K, respectively, in excellent agreement with the theoretical prediction from Section \ref {sec:floats}, and in close correspondence with the onset of the fastest acceleration in the flame speed. After the onset of detonation, the flame speed accelerates beyond the Chapman-Jouguet detonation velocity $D_{\rm CJ}$, corresponding to the case of an overdriven detonation \citep {khokhlov89}. The overdriven detonation speed here is the result of multiple nearly-simultaneous detonations arising within the computational volume, which arise as $L / r_{\rm crit} > 1$. The fuel is quickly expended on a detonation crossing time $L / D_{\rm CJ} \simeq 0.01$ s, and the burning rate relaxes down to $\sim 2 \times 10^7$ cm/s, which is below the Chapman-Jouguet subsonic deflagration speed $S_{\rm CJ} = 4.2 \times 10^7$ cm/s in the ash.

The detonation rapidly consumes both $^{4}$He  and $^{12}$C, as their abundances  drop rapidly to 16.6\% and 0\%, respectively.  This final unburned  $^{4}$He fuel is expected, as the initial $^{12}$C abundance just prior to detonation is less than the critical amount (21\% by mass) required for complete $^4$He burning by $\alpha$ captures \citep {gronowetal20}. The remaining final composition of the simulation is primarily $^{56}$Ni, though this is an artifact of the extended burn facilitated by the effectively infinite spatial domain of the fully periodic boundaries. In a more realistic open geometry, we expect that helium detonations at these densities would lead to incomplete burning and the production of significant quantities of $^{44}$Ti, as previous studies have found \citep {timmesetal96, Holcomb_2013}.

The simulation LowDen was advanced to $7.38 \times 10^8$ K and 309 ms.
Detonation did not arise in this simulation, consistent with the predictions of Section \ref {sec:floats}.
Table 1 lists the results from the runs, including both estimates for the detonation temperatures.

\begin{table}[ht!]
\caption{Results from the runs. Mass-weighted mean temperature at the time of detonation are listed for both the maximal $^{12}$C criterion as well as the Chapman-Jouguet criterion (see text).}
  \begin{center}
      \begin{tabular}{|c|c|c|c|}
        \hline
	      Model & Density (g cm$^{-3}$)  & $T_{\rm C12}$(K) & $T_{\rm CJ}$(K) \\
        \hline\hline
	      LowDen & $10^5$  & None & None  \\
	      HighDen & $10^6$  & $1.44 \times 10^9$ & $1.78 \times 10^9$ \\
        \hline
   \end{tabular}
  \end{center}
  \label{runs_summary}
\end{table}

\section{Discussion and Conclusion}\label{sec:discussion}

This paper establishes how helium detonation may arise in realistic turbulent astrophysical environments in the distributed burning regime. In particular, a novel pathway for the detonation of even initially pure helium was demonstrated owing to the nucleosynthesis of $^{12}$C seed nuclei by 3$\alpha$, facilitated by strong turbulent dissipation.

Our key conditions for detonation initiation, equations \ref {distburn} and \ref {rcrit} are essentially the same as equation 19 of \citet {woosley07}. Woosley notes correctly that thermodynamic properties  ``can vary greatly depending on the instantaneous local values in a given flame" and favored the burning rate calculated within an isobaric mixture over this expression. The burning rate generally exhibits overwhelmingly the greatest sensitivity to temperature in equation \ref {distburn}. \citet {zenatifisher23} showed that the effect of Gaussian fluctuations in homogeneous isotropic turbulence is to increase the net burning rate by a factor, dependent upon the RMS temperature fluctuations on the integral scale, which they calculate exactly. However, the specific 3-$\alpha$ nuclear burning rates within the temperature range close to detonation of $T \sim 10^9$ K is not highly sensitive to temperature (scaling approximately as $\propto$ T). Even with the addition of $^{12}$C seed nuclei, the nuclear burning rate scales as $T^8 - T^{10}$, and the resulting enhancement to the burning rate is negligible \citep [see figure 1 in][]{zenatifisher23}.  We infer that equations \ref {distburn} and \ref {rcrit} are valid for the case of pure $^{4}$He burning and modest admixtures of $^{12}$C in the temperature close to $10^9$ K considered in this paper.

Further work is needed to more broadly explore the parameters for turbulent mechanism, including the possible impact of proton-catalyzed $\alpha$ captures such as $^{12}$C (p, $\gamma$) $^{13}$N($\alpha$, p)$^{16}$O, which may  enhance the reaction rates and the detonation initiation conditions from the models computed here based upon a 19 isotope network \citep {shenmoore14}. Approximate networks, such as the subch.simple network of \citet {chenetal23}, which include this proton-catalyzed pathway with just 22 isotopes, offer a promising route to efficiently capture these critical nuclear reactions for helium burning with a minimum of computational expense. 
The turbulent models can ultimately help provide subgrid models or criteria in global simulations to determine under what conditions the full WD can be detonated, and tested directly both against observations of SNe Ia transients as well as surviving hypervelocity Gaia donors.

\begin {acknowledgements}

This material is based upon work supported by the U.S. Department of Energy, Office of Science, Office of Advanced Scientific Computing Research, Department of Energy Computational Science Graduate Fellowship under Award Number DE-SC0020347.  The authors acknowledge insightful conversations with Fritz R\"opke, Javier Mor\'an Fraile, Hagai Perets, and Noam Soker. R.T.F. acknowledges support from NASA ATP awards 80NSSC18K1013 and 80NSSC22K0630, NASA XMM-Newton award 80NSSC19K0601, and NASA HST-GO-15693.

R.T.F. also gratefully acknowledges support from the DAAD, the Institute for Advanced Studies and the Heidelberg Institute for Theoretical Studies (HITS). This work used the Extreme Science and Engineering Discovery Environment (XSEDE) Stampede 2 supercomputer at the University of Texas at Austin’s Texas Advanced Computing Center through allocation TG-AST100038. XSEDE is supported by National Science Foundation grant number ACI-1548562 \citep{townsetal14}.

\end {acknowledgements}

\software {FLASH 4.0.1 \citep{fryxell2000flash, dubeyetal13}, yt  \citep{Turk_2011}, Python programming language \citep{vanrossumetal1991}, Numpy \citep{vanderwaltetal2011}, IPython \citep{perezetal2007}, Matplotlib \citep{hunter2007}}.

\bibliography{turbulent_helium}{}
\bibliographystyle{aasjournal}

\end{document}